%% file: cipanp_leskovec.tex
\documentclass[12pt]{article}
\usepackage[utf8]{inputenc}
\usepackage{textcomp}
\usepackage{graphicx}
\usepackage{color}
\usepackage{longtable}
\usepackage[breaklinks=true]{hyperref}
\usepackage{latexsym}
\usepackage{amsmath}
\usepackage{amssymb}
\usepackage{dsfont}
\usepackage{verbatim}
\usepackage{bm}
\usepackage{bbm}
\usepackage{placeins}
\usepackage{newunicodechar}
\usepackage{floatrow}
\usepackage{cite}

\inputencoding{utf8}
\newunicodechar{γ}{$\gamma$}
\newunicodechar{ρ}{$\rho$}
\newunicodechar{π}{$\pi$}
\newunicodechar{Δ}{$\Delta$}
\newunicodechar{η}{$\eta$}
\newunicodechar{ψ}{$\psi$}

\newfloatcommand{capbtabbox}{table}[][\FBwidth]

\newcommand\pubnumber{CIPANP2018-Leskovec}
\newcommand\pubdate{\today}

\def\napoli{Department of Physics\\
University of Arizona, AZ-85721, USA}
\def\support{}

\def\Title#1{\begin{center} {\Large #1 } \end{center}}
\def\Author#1{\begin{center}{ \sc #1} \end{center}}
\def\Address#1{\begin{center}{ \it #1} \end{center}}

\newcommand\pubblock{\rightline{\begin{tabular}{l} \pubnumber\\
         \pubdate  \end{tabular}}}
\newenvironment{Abstract}{\begin{quotation}  }{\end{quotation}}
\newenvironment{Presented}{\begin{quotation} \begin{center} 
             PRESENTED AT\end{center}\bigskip 
      \begin{center}\begin{large}}{\end{large}\end{center} \end{quotation}}
\def\Acknowledgements{\bigskip  \bigskip \begin{center} \begin{large}
             \bf ACKNOWLEDGEMENTS \end{large}\end{center}}
\input econfmacros.tex
\begin{document}
\begin{titlepage}
\pubblock

\vfill
\Title{A Lattice QCD study of the $\rho$ resonance}
\vfill
\Author{ Luka Leskovec\\
{\small in collaboration with:\\
Constantia Alexandrou, Stefan Meinel, John Negele, Srijit Paul, Marcus Petschlies, Andrew Pochinsky, Gumaro Rendon, and Sergey Syritsyn
} \support}
\Address{\napoli}
\vfill
\begin{Abstract}
We present a lattice QCD study of the $\rho$ resonance with $N_f=2+1$ clover fermions at a pion mass of approximately $320$ MeV and lattice size $3.6$ fm. We consider two processes involving the $\rho$. The first process is elastic scattering of two pions in P-wave with isospin $1$. Using the L\"uscher method we determine the scattering phase shift, from which we obtain the $\rho$ resonance mass and decay width $\Gamma(\rho\to\pi\pi)$. The second process is the radiative transition $\pi\gamma\to\pi\pi$, where we follow the Brice\~no-Hansen-Walker-Loud approach to determine the transition amplitude in the invariant mass region near the $\rho$ resonance and for both space- and time-like photon momentum. This allows us to determine the coupling between the $\rho$, the pion and the photon, and the resulting $\rho$ radiative decay width.
\end{Abstract}
\vfill
\begin{Presented}
Thirteenth Conference on the Intersections of Particle and Nuclear Physics\\
Palm Springs CA, USA,  May 29 -- June 3, 2018
\end{Presented}
\vfill
\end{titlepage}
\def\thefootnote{\fnsymbol{footnote}}
\setcounter{footnote}{0}
%

%%%%%%%%%%%%%%%%%%%%%%%%%%%%%%%%%%%%%%%%%%%%%%%%%%%%%%%%%%%%%%%%%%%%%%%%%%%%%%%%%%%%%%%%%%%%%%%%%%%%%%%%%%%
\section{Introduction}\label{sec_introduction}
%%%%%%%%%%%%%%%%%%%%%%%%%%%%%%%%%%%%%%%%%%%%%%%%%%%%%%%%%%%%%%%%%%%%%%%%%%%%%%%%%%%%%%%%%%%%%%%%%%%%%%%%%%%

The spectrum of hadrons is a jungle of particles arising from the strong interactions between quarks and gluons. We differentiate between stable hadrons, i.e. those that cannot decay via the strong interactions, and unstable hadrons, i.e. those that can. The simplest example of an unstable hadron is the $\rho$ meson, which is an isotriplet with quantum numbers $J^{PC}=1^{--}$. While the $\rho$ couples to multiple decay modes, for example $K\bar{K}$, $\pi\pi\pi\pi$, $\pi \pi$ and $\pi \gamma$, we focus only on the latter two in this report. This simplification can be made because we perform our lattice QCD calculation at light quark masses corresponding to $m_{\pi}\approx 320$ MeV, where the $K\bar{K}$ and $\pi\pi\pi\pi$ thresholds are above the energy region we are focusing on. The coupling of the  $\rho$ to the  $\pi \pi$ channel was previously investigated with lattice QCD in Refs. \cite{Aoki:2007rd,Feng:2010es,Lang:2011mn,Aoki:2011yj,Pelissier:2012pi,Dudek:2012xn,Wilson:2015dqa,Bali:2015gji,Bulava:2016mks,Hu:2016shf,Guo:2016zos,Fu:2016itp,Alexandrou:2017mpi,Andersen:2018mau},
%%check rho references if they are all
and the coupling to $\pi \gamma$ in Refs.~\cite{Briceno:2015dca,Briceno:2016kkp,Alexandrou:2018jbt}.

To determine the coupling of the $\rho$ resonance to two pions and thus establish its strong decay width, we make use of the L\"uscher formalism, which relates the finite-volume spectrum in various moving frames and irreducible representations with the infinite-volume scattering matrix \cite{Luscher:1990ux,Briceno:2017max}.
%% add the appropriate refs
The appropriate formalism to handle radiative transitions of multi-hadron states, which we employ here, is the generalization of Lellouch and L\"uscher's work in Ref.~\cite{Lellouch:2000pv} by Brice\~no, Hansen and Walker-Loud (BHWL) \cite{Briceno:2014uqa,Briceno:2015csa}. 

%%%%%%%%%%%%%%%%%%%%%%%%%%%%%%%%%%%%%%%%%%%%%%%%%%%%%%%%%%%%%%%%%%%%%%%%%%%%%%%%%%%%%%%%%%%%%%%%%%%%%%%%%%%
\section{Lattice Setup}\label{sec_setup}
%%%%%%%%%%%%%%%%%%%%%%%%%%%%%%%%%%%%%%%%%%%%%%%%%%%%%%%%%%%%%%%%%%%%%%%%%%%%%%%%%%%%%%%%%%%%%%%%%%%%%%%%%%%

We use a single lattice gauge-field ensemble with light quark masses corresponding to $m_\pi\approx 320$ MeV and a strange-quark mass consistent with its physical value. The number of lattice points is $N_L^3 \times N_t = 32^3 \times 96$. The lattice spacing $a$, determined from the $\Upsilon(2S)-\Upsilon(1S)$ splitting calculated with NRQCD, is equal to $0.11403(77)$ fm, leading to a physical spatial volume of approximately $(3.6$ fm$)^3$. 

%%%%%%%%%%%%%%%%%%%%%%%%%%%%%%%%%%%%%%%%%%%%%%%%%%%%%%%%%%%%%%%%%%%%%%%%%%%%%%%%%%%%%%%%%%%%%%%%%%%%%%%%%%%
\section{\texorpdfstring{$P$-wave $\pi\pi$ scattering in $I=1$}{}}\label{sec_rhopipi}
%%%%%%%%%%%%%%%%%%%%%%%%%%%%%%%%%%%%%%%%%%%%%%%%%%%%%%%%%%%%%%%%%%%%%%%%%%%%%%%%%%%%%%%%%%%%%%%%%%%%%%%%%%%

We calculated the lattice spectra in several moving frames and irreducible representations \cite{Alexandrou:2017mpi} and determined the scattering phase shifts using the L\"uscher method. To describe the phase shifts we use the Breit-Wigner formula
\begin{align}
\label{eq:ps}
\delta_1(s) = \arctan \frac{\sqrt{s}\,\Gamma(s)}{m_R^2 - s},
\end{align}
where $s$ is the invariant mass, $\delta_1(s)$ is the $P$-wave scattering phase shift, $m_R$ is the $\rho$ resonance mass and $\Gamma(s)$ is the decay width. We investigate two different parametrizations: 
\begin{itemize}
  \item {\bf BW I:} $P$-wave decay width:
  \vspace{-0.3cm}
  \begin{align}
  \label{eq:Gamma_Pwave}
  \Gamma_{I}(s) = \frac{g_{\rho\pi\pi}^2}{6\pi} \frac{k^{3}}{s},
  \end{align}
  where $g_{\rho\pi\pi}$ is the coupling between the resonance $\rho$ and the $\pi\pi$ scattering channel and $k$ is the scattering momentum, $\sqrt{s} = 2\sqrt{m_{\pi}^2 + k^{2}}$, and
  \item {\bf BW II:} $P$-wave decay width modified with Blatt-Weisskopf barrier factors \cite{VonHippel:1972fg}:
  \vspace{-0.3cm}
  \begin{align}
  \label{eq:Gamma_PwaveBW}
  \Gamma_{II}(s) = \frac{g_{\rho\pi\pi}^2}{6\pi} \frac{k^{3}}{s}\: \frac{1 + (k_R r_0)^2}{1 + (k r_0)^2},
  \end{align}
  where $k_R$ is the scattering momentum at $\sqrt{s}=m_R$ and $r_0$ is the centrifugal barrier radius.
\end{itemize} 
\begin{figure}
\begin{floatrow}
\ffigbox{%
\includegraphics[width=0.48\textwidth]{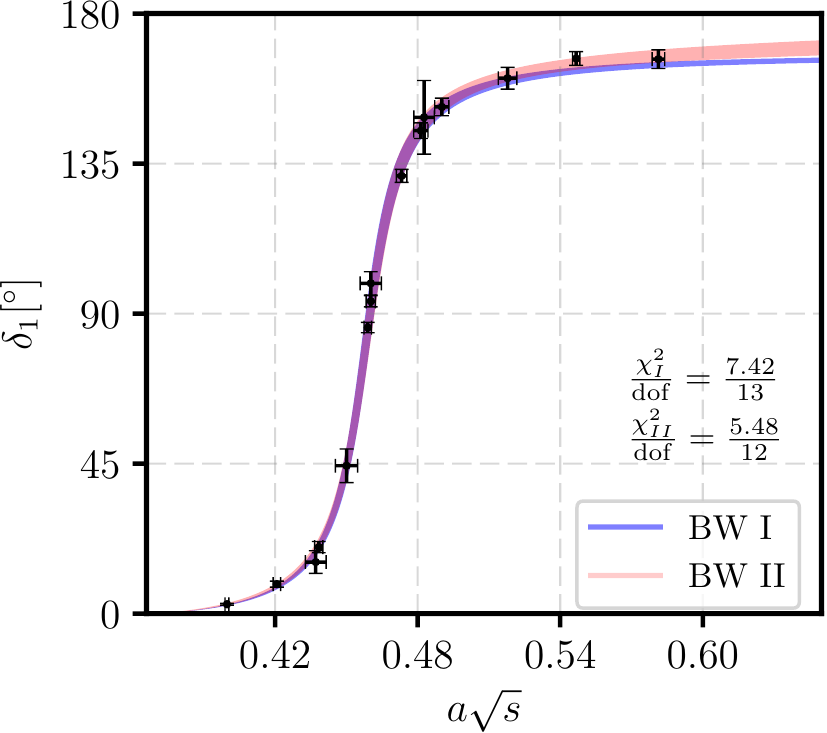}
}{%
\caption{\label{fig:delta_rho} $P$-wave $\pi\pi$ scattering phase shift in $I=1$ as a function of invariant mass.}
}
\capbtabbox{%
\begin{tabular}{| l | c | c |}
\hline
Model                      & {\bf BW I}    & {\bf BW II}   \cr
\hline
$\frac{\chi^2}{{\rm dof}}$ &   $0.571$     &  $0.457$      \cr 
$am_R$                     & $0.4599(19)$  &  $0.4600(18)$ \cr
$g_{\rho\pi\pi}$           & $5.76(16)$    &  $5.79(16)$   \cr    
$(r_0/a)^2$                &    --          &  $8.6(8.0)$   \cr
\hline
\end{tabular}
\vspace{2.5cm}
}{%
\caption{\label{tab:rho} Parameters of the Breit-Wigner parametrizations {\bf BW I} and {\bf BW II} obtained from fitting the discrete phase shift points with statistical uncertainties only \cite{Alexandrou:2017mpi}.}
}
\end{floatrow}
\end{figure}
The resulting fits for both parametrizations are shown in Fig.~\ref{fig:delta_rho}, where the blue line corresponds to {\bf BW I} and the red line to {\bf BW II}. The numerical results for the parameters $m_R$, $g_{\rho\pi\pi}$ and $r_0$ are listed in Table \ref{tab:rho}. There are only minor and statistically not very significant differences between the two parametrizations that appear in the high-energy region $a\sqrt{s}\geq 0.55$, where {\bf BW II} describes the data slightly better \cite{Alexandrou:2017mpi}. Overall, we find that both parametrizations describe our results for the elastic $I=1$ $\pi\pi$ $P$-wave scattering well.

%%%%%%%%%%%%%%%%%%%%%%%%%%%%%%%%%%%%%%%%%%%%%%%%%%%%%%%%%%%%%%%%%%%%%%%%%%%%%%%%%%%%%%%%%%%%%%%%%%%%%%%%%%%
\section{\texorpdfstring{$\pi \gamma \to \pi\pi$ transition amplitude}{}}\label{sec_rhopigamma}
%%%%%%%%%%%%%%%%%%%%%%%%%%%%%%%%%%%%%%%%%%%%%%%%%%%%%%%%%%%%%%%%%%%%%%%%%%%%%%%%%%%%%%%%%%%%%%%%%%%%%%%%%%%

Because the $\rho$ is not a QCD asymptotic state, but rather a resonance in $P$-wave $\pi\pi$ scattering with $I=1$, the observables related to the resonance photoproduction processes $\pi \gamma \to \rho$ are obtained from the more general process $\pi\gamma \to \pi \pi$. This process is described by the transition amplitude $V_{\pi\gamma\to\pi\pi}$, which is a function of both the photon four-momentum transfer $q^2$ and the $\pi\pi$ invariant mass $s$. The transition $\pi\gamma \to \rho$ can however be defined by analytical continuation to the $\rho$ pole located at $s_P=m_R^2 + i \Gamma_R m_R$. 
\begin{figure}
\includegraphics[width=0.65\textwidth]{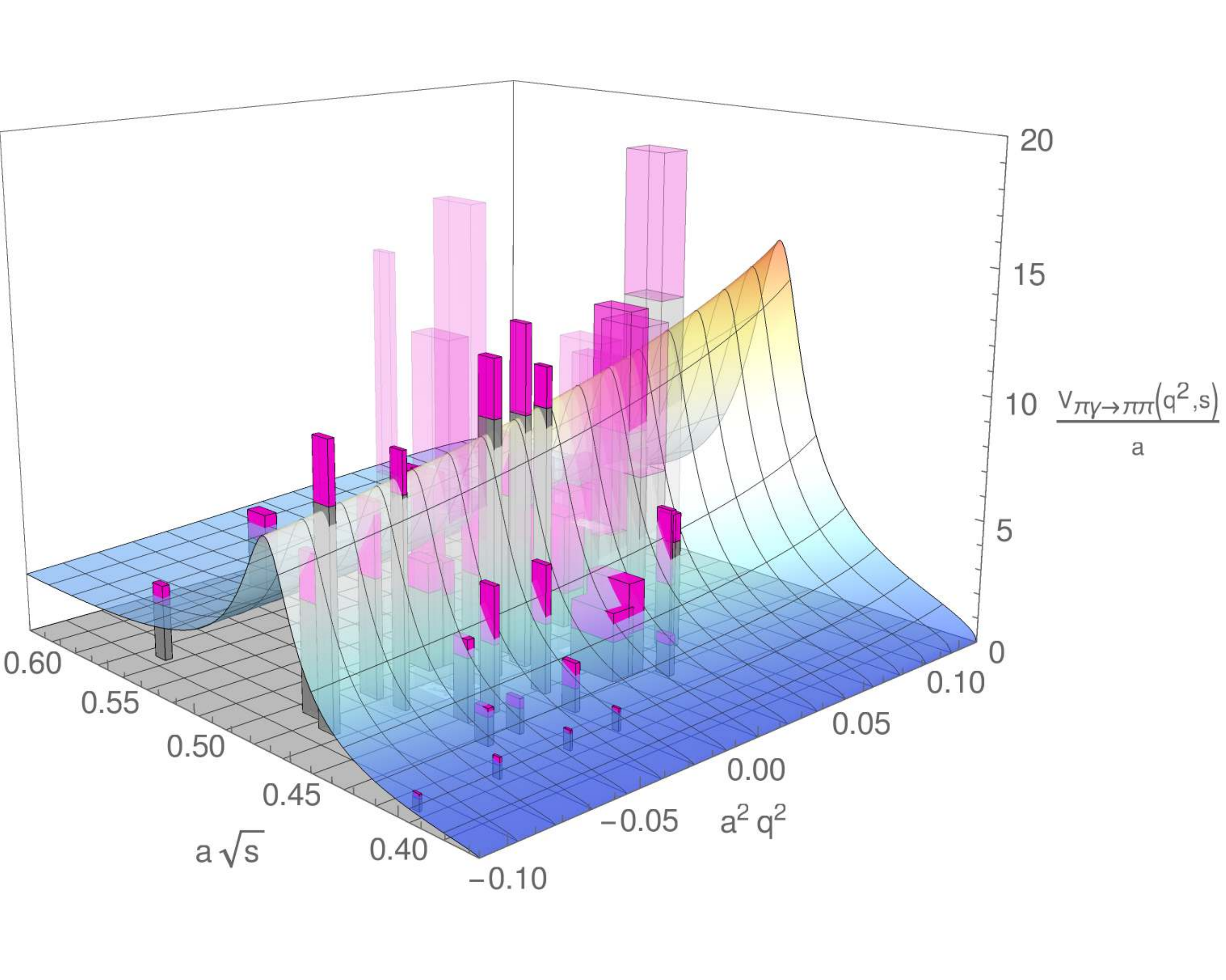}
\caption{\label{fig:3d} 3-D plot of the transition amplitude $V_{\pi\gamma\to\pi\pi} $. Lattice QCD results are shown as the vertical bars, where the widths and depths correspond to the uncertainties in $a\sqrt{s}$ and $(aq)^2$, and the magenta sections at the tops cover the range from $V_{\pi\gamma\to\pi\pi}-\sigma_{V_{\pi\gamma\to\pi\pi}}$ to $V_{\pi\gamma\to\pi\pi}+\sigma_{V_{\pi\gamma\to\pi\pi}}$. For clarity, the opacity is reduced in data points with larger uncertainties. The surface shows the central value of the chosen parametrization ``BWII F1 K2''.
}
\end{figure}
The transition amplitude $V_{\pi\gamma\to\pi\pi}$ is associated with the infinite volume matrix element $\langle \pi\pi | J^\mu(0) | \pi \rangle$ by
\begin{align}
\label{eq:LLdecomp}
\langle \pi\pi | J^\mu(0) | \pi \rangle = \frac{2i V_{\pi\gamma\to\pi\pi}(q^2,s)}{m_\pi} \epsilon^{\nu\mu\alpha\beta} \epsilon_{\nu}(P,m) (p_\pi)_\alpha P_\beta,
\end{align}
where $q=p_\pi-P$ is the photon four-momentum transfer. The transition amplitude $V_{\pi\gamma\to\pi\pi}$ has a manifest pole at $s=s_P$, and can be written with the help of Watson's theorem as
\begin{align}
\label{eq:V_BW_d}
V_{\pi\gamma\to\pi\pi}(q^2,s) & =  \sqrt{\frac{16 \pi s \Gamma(s)}{k}} \frac{F(q^2,s) }{m_R^2 - s - i\sqrt{s}\Gamma(s)}.
\end{align}

The form factor $F(q^2,s)$ is thus free of poles in $s$ within the energy region of interest, $a\sqrt{s}<0.6$. Following a general Taylor-expansion approach we parametrize $F(q^2,s)$ using 
 \begin{align}
 \label{eq:F}
 F(q^2,s) = \frac{1}{1 - \frac{q^2}{m_P^2}} \sum_{n,m} A_{nm} z^n \mathcal{S}^m,
 \end{align}
where $m_P$ is the pole in the $t$-channel, and the two variables $\mathcal{S}$ and $z$ are defined as \cite{Boyd:1994tt,Boyd:1997qw,Bourrely:2008za}:
\begin{align}
\mathcal{S}&=\frac{s-m_R^2}{m_R^2} \\
z &= \frac{\sqrt{t_+ - q^2} - \sqrt{t_+}}{\sqrt{t_+ - q^2} + \sqrt{t_+}}, \text{ where }t_+ = (2 m_\pi)^2.
\end{align}
We consider three different families of truncations of the series in Eq.~\ref{eq:F} leading to several parametrizations of the transition amplitude $V_{\pi\gamma\to\pi\pi}$ \cite{Alexandrou:2018jbt}. A 3-D representation of the transition amplitude $V_{\pi\gamma\to\pi\pi}$ is shown for the chosen parametrization ``BW II F1 K2'' in Fig.~\ref{fig:3d}.

In practice, the resonant form factor is determined by evaluating $F$ at the $\rho$ pole:
\begin{equation}
 F_{\pi\gamma\to\rho}(q^2) = F(q^2,\, m_R^2 + i m_R \Gamma_R). \label{eq:Fresonant}
\end{equation}
It becomes equal to the photocoupling $G_{\rho\pi\gamma}$ at zero momentum transfer, $q^2=0$. The physical observable we consider is the $\rho$ radiative decay width $\Gamma(\rho\to\pi\gamma)$ determined by the photocoupling $G_{\rho\pi\gamma}=F(0, m_R^2 + i m_R \Gamma_R)$,
\begin{align}
&\Gamma(\rho \to \pi \gamma) =\frac{2}{3}\alpha \left(\frac{m_R^2 - m_\pi^2}{2m_R}\right)^3  \frac{|G_{\rho\pi\gamma}|^2}{m_\pi^2}. \label{eq:radwidth}
\end{align}

\begin{figure}
\includegraphics[width=0.48\textwidth]{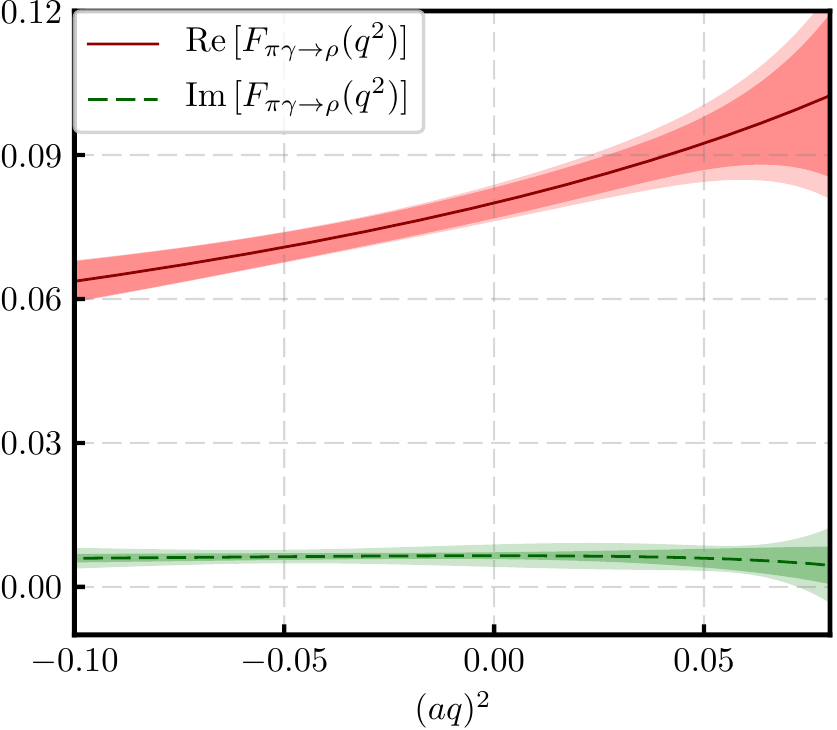}%
\includegraphics[width=0.48\textwidth]{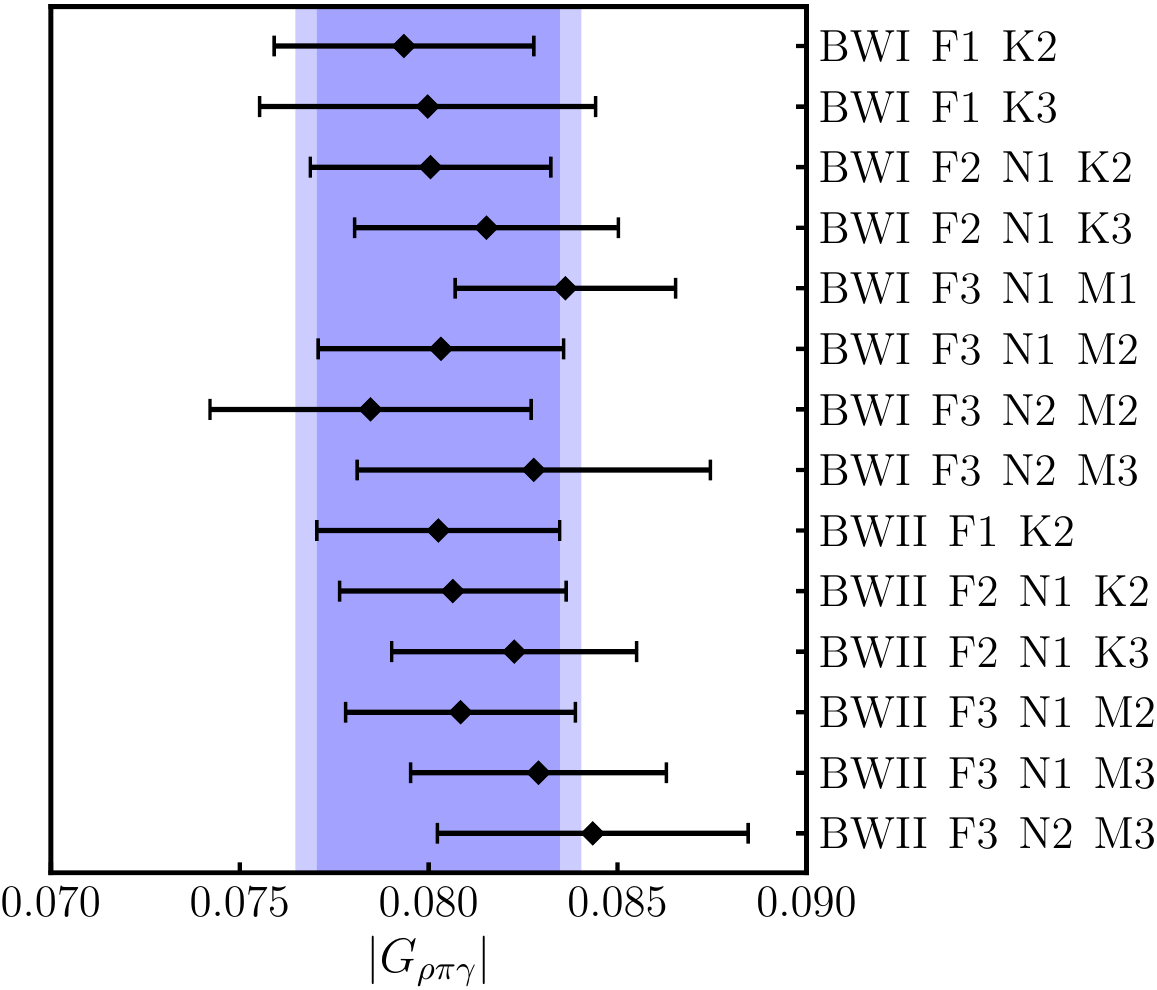}
\caption{\label{fig:PC} Left panel: Resonant transition form factor $F_{\pi\gamma\to\rho}$. The inner shaded region represents the statistical and systematical uncertainties, and the outer shaded region represents  the parametrization uncertainty. Right panel: Value of the photocoupling for the parametrizations under consideration. The data-point errorbars are associated with lattice statistical and systematical uncertainties. The inner shaded region corresponds to the errorbar of the chosen model, ``BW II F1 K2'', while the outer shaded region represents the parametrization uncertainty.}
\end{figure}

The resonant transition form factor is shown in the left panel of Fig.~\ref{fig:PC}, where the inner shaded region represents the statistical and systematical uncertainties determined on the lattice. The outer shaded region indicates the parametrization uncertainty. The photocouplings determined for each of the parametrizations are shown in the right panel of Fig.~\ref{fig:PC}; we find the photocoupling at $m_\pi\approx320$ MeV to be
\begin{align}
|G_{\rho\pi\gamma}| = 0.0802(32)(20).
\end{align}
Due to the larger-than physical pion mass in our calculation the thresholds are much closer to the resonances than in nature. By assuming that the photocoupling is a pion-mass-independent quantity, we use the physical values of the pion masses and $\rho$ mass to determine the radiative decay width to be $\Gamma(\rho\to\pi\gamma) \,=\, 84.2(6.7)(4.3)\, {\rm keV}$. The number in the first bracket is the combined statistical and systematic uncertainty and the number in the second bracket is the parametrization uncertainty.

%%%%%%%%%%%%%%%%%%%%%%%%%%%%%%%%%%%%%%%%%%%%%%%%%%%%%%%%%%%%%%%%%%%%%%%%%%%%%%%%%%%%%%%%%%%%%%%%%%%%%%%%%%%
\section{Summary}\label{sec_summary}
%%%%%%%%%%%%%%%%%%%%%%%%%%%%%%%%%%%%%%%%%%%%%%%%%%%%%%%%%%%%%%%%%%%%%%%%%%%%%%%%%%%%%%%%%%%%%%%%%%%%%%%%%%%
We have presented results of two of our recent lattice QCD studies, the determination of the strong decay width of the $\rho$ resonance and the calculation of the $\rho$ resonance radiative decay width. While our calculation was performed at a non-physical value of the quark mass, the couplings $g_{\rho\pi\pi}$ and $G_{\rho\pi\gamma}$ are already close to their physical values. Future studies are needed to perform chiral extrapolations to the physical point \cite{Bolton:2015psa,Hu:2017wli,Bruns:2017gix} where direct comparisons to experiment can be made.\\

\vspace{0.5cm}

\Acknowledgements
We are grateful to Kostas Orginos for providing the gauge field ensemble, which was generated using resources provided by XSEDE (supported by National Science  Foundation Grant No.~ACI-1053575). We thank R. A. Brice\~no, M. Hansen, and C. B. Lang for valuable discussions.
SM and GR were supported in part by National Science Foundation Grant No.~PHY-1520996; SM, GR, and LL were also supported in part by the U.S.~Department of Energy Office of High Energy Physics under Grant No.~DE-{}SC0009913. SM and SS futher acknowledge support by the RHIC Physics Fellow Program of the RIKEN BNL Research Center. JN and AP were supported in part by the U.S. Department of Energy Office of Nuclear Physics under Grant Nos.~DE{}-SC-0{}011090 and DE-{}FC02-06ER41444. We acknowledge funding from the  European Union's Horizon 2020 research and innovation programme under the Marie Sklodowska-Curie grant agreement No 642069. SP is a Marie Sklodowska-Curie  fellow supported by the HPC-LEAP joint doctorate program. This research used resources of the National Energy Research Scientific Computing Center (NERSC), a U.S. Department of Energy Office of Science User Facility operated under Contract No. DE-{}AC02-05CH11231. The computations were performed using the Qlua software suite \cite{QLUA}.

{\small
%\bibliographystyle{utphys-noitalics}
%\bibliography{cipanp_leskovec} 
\providecommand{\href}[2]{#2}\begingroup\raggedright\endgroup
}

\end{document}

%% file: econfmacros.tex
\def\beq{\begin{equation}}
\def\eeq#1{\label{#1}\end{equation}}
\def\eeqn{\end{equation}}

\def\beqa{\begin{eqnarray}}
\def\eeqa#1{\label{#1}\end{eqnarray}}
\def\eeqan{\end{eqnarray}}

\let\bar=\overbar

\def\Dslash{\not{\hbox{\kern-4pt $D$}}}
\def\dslash{\not{\hbox{\kern-2pt $\del$}}}

\def\msb{{\bar{\ssstyle M \kern -1pt S}}}